\documentclass[twocolumn,showpacs,showkeys,preprintnumbers]{revtex4-1}
\usepackage{amssymb}

\usepackage{amsfonts}

\usepackage{latexsym}

\usepackage{dcolumn}

\usepackage{amsmath}

\usepackage{graphicx}

\usepackage{psfrag,pstricks}

\newcommand{\eqn}{\ref}

\begin{document}

\title{Nonlinear effects of atomic collisions on the optomechanical properties of a Bose-Einstein condensate in an optical cavity}

\author{A. Dalafi$^{1}$ }
\email{adalafi@yahoo.co.uk}

\author{M. H. Naderi$^{1,2}$}
\author{ M. Soltanolkotabi$^{1,2}$}
\author{ Sh. Barzanjeh$^{1,3}$}
\affiliation{$^{1}$ Department of Physics, Faculty of Science, University of Isfahan, Hezar Jerib, 81746-73441, Isfahan, Iran\\
$^{2}$Quantum Optics Group, Department of Physics, Faculty of Science, University of Isfahan, Hezar Jerib, 81746-73441, Isfahan, Iran\\
$^{3}$ School of Science and Technology, Physics Division, Universita di Camerino, I-62032 Camerino (MC), Italy}

\date{\today}

\begin{abstract}

In this paper, we have investigated theoretically the influence of atomic collisions on the behaviour of a  one-dimensional Bose-Einstein condensate inside a driven optical cavity. We develop the discrete-mode approximation for the condensate taking into account the interband transitions due to the s-wave scattering interaction. We show that in the Bogoliubov approximation the atom-atom interaction shifts the energies of the excited modes and also plays the role of an optical parametric amplifier for the Bogoliubov side mode which can affect its normal-mode splitting behaviour. On the other hand due to the atomic collisions the resonance frequency of the cavity is shifted which leads to the decrease of the number of cavity photons and the depletion of the Bogoliubov mode. Besides, it reduces the effective atom-photon coupling parameter which consequently leads to the decrease of the entanglement between the Bogoliubov mode and the optical field. 
\end{abstract}


\maketitle

\section{Introduction}
%
%
The emerging field of optomechanics concerns with the study of the mechanical effects of light on mesoscopic and macroscopic mechanical oscillators. This phenomenon has been realized in the optomechanical systems consisting of  an optical cavity with a movable end-mirror or with a membrane-in-the middle. The radiation pressure exerted by the light inside the optical cavity couples the moving mirror or the membrane which acts as a mechanical oscillator to the optical field. This optomechanical coupling has been employed for a wide range of applications such as the cavity cooling of microlevers and nanomechanical resonators to their quantum mechanical ground state \cite{Genes 2008, Teufel, Liberato, Ludwig, Barzanjeh2}, producing high precision detectors for measuring weak forces and small displacements and also providing a good approach for fundamental studies of the transition between the quantum and the classical world \cite{Bradaschia,LaHaye,Kippenberg}.
 
On the other hand an alternative path to the studies of cavity optomechanics has been provided experimentally by systems consisting of ultracold atomic ensembles trapped in optical cavities \cite{Brenn Nature,Gupta,Brenn Science,Ritter Appl. Phys. B} where the excitation of a collective mode of the cold gas plays the role of the vibrational mode of the mirror. The standing electromagnetic wave inside the cavity forms a periodic potential, the so-called optical lattice, in which the cold atoms exhibit phenomena typical of solid state physics like the formation of energy bands and Bloch oscillations \cite{Morsch}. 

In such systems with high finesse cavities the atom-light interaction is enhanced because the atoms are collectively coupled to the same optical mode. Besides, in the dispersive regime where the laser pump is far detuned from the atomic resonance the excited electronic state of the atoms can be adiabatically eliminated and consequently the only degrees of freedom of atoms will be their mechanical motions \cite{Masch Ritch 2004, Dom Ritch 2003, Masch Ritch 2005}. For low photon numbers when the optical grating produced by the intracavity optical field is very shallow, one can approximately restrict the dynamics to the first two motional modes \cite{Kanamoto 2010, Nagy Ritsch 2009}. In more recent theoretical investigations \cite{Konya Domokos 2011, Zhang 2009} it has been shown that the simple two-mode model of a Bose-Einstein condensate (BEC) can be improved by considering higher motional modes.

In spite of similarities between the two kinds of optomechanical systems (with a moving mirror and with a BEC), there are some essential differences between them. Firstly, in contrast to the moving mirror of the optomechanical sytems which is placed in the harmonic potential of its spring, the excitation modes of the BEC are not based on the presence of such an external harmonic potential \cite{Brenn Science}. Secondly, their parameters are realized in different regimes; the frequency of the excitation mode of the BEC, i.e., recoil frequency, is well below the dipole coupling strength while the oscillation frequency of a moving mirror is of the same order of magnitude as the coupling strength \cite{Ritter Appl. Phys. B,Nagy Ritsch 2009}. On the other hand in addition to pumping the cavity from one of the end mirrors, in the cavities equipped with atomic gas it is also possible to pump the atoms from the side of the cavity\cite{Maschler2008}.

One of the most important characteristics of the many-body systems is the two-body collision which can affect the properties of the system. So in order to study the dynamics of a BEC gas in a realistic experimental situation, it is necessary to take it into account. In the optomechanical systems containing the atomic gas there are two kinds of atom-atom interactions which are the origins of the system nonlinearities. Firstly, due to the atom-photon interaction the potential acting on the condensate depends in a highly nonlocal and nonlinear way on the condensate itself which leads to the long range atomic interaction mediated by the cavity field \cite{Asboth,Zhang 2009}. On the other hand there is an intrinsic nonlinearity due to the s-wave scattering which can take place at arbitrary momentum values and causes a broadening of the atomic momentum distribution \cite{Goldbaum Meystre arXiv} due to the intraband transitions and also can scatter atoms to the other bands (interband transitions)\cite{Szirmai 2010,Morsch}. Furthermore, there is another kind of nonlinearitiy due to the finiteness of the particle number of the system which can be neglected in the thermodynamic limit where the total number of atoms is very large \cite{Gardiner97}.

 In this work, we are going to extend the two-mode model considering the effects of atom-atom interaction for a one-dimensional BEC in an optical cavity. We consider the system in the low photon regime but will not restrict our treatment to the weak atom-atom interaction. In fact the atomic collisions can populate several nonzero quasimomenta of the energy bands (intraband transition) and can also cause the atoms to be scattered to the other bands (interband transitions). In order to have a simplified optomechanical model we take into account just the lowest nozero quasimomenta in the first Brillouin zone . We show that in the Bogoliubov approximation the atom-atom interaction not only shifts the energies of the excited modes but also play the role of an oprtical parametric amplifier (OPA) for the Bogoliubov side mode which can affect its normal-mode splitting behaviour \cite{Agarwal 2009}. On the other hand due to the atomic collisions the resonance frequency of the cavity is shifted which leads to the decrease of the number of cavity photons and the depletion of the Bogoliubov mode. These results are in good agreement with those obtained by the numerical solutions of the full description of Gross-Pitaevskii Equation (GPE) \cite{Zhang 2009, Horak 2000}.

The paper is structured as follows. In Sec. II we will first give a thorough theoretical description of the many-body system under consideration and then derive a simplified optomechanical model. In Sec. III the quantum Langevin equations (QLEs) are derived and linearized around the semiclassical steady state. In Sec. IV we study the mean-field solutions and fluctuations of the system and in Sec. V we derive the spectrum of the Bogoliubov mode and investigate the normal mode splitting (NMS). Finally, our conclusions are summarised in Sec. VI.
\section{Theoretical Description of the System}

\begin{figure}[ht]
\centering
\includegraphics[width=3.5 in]{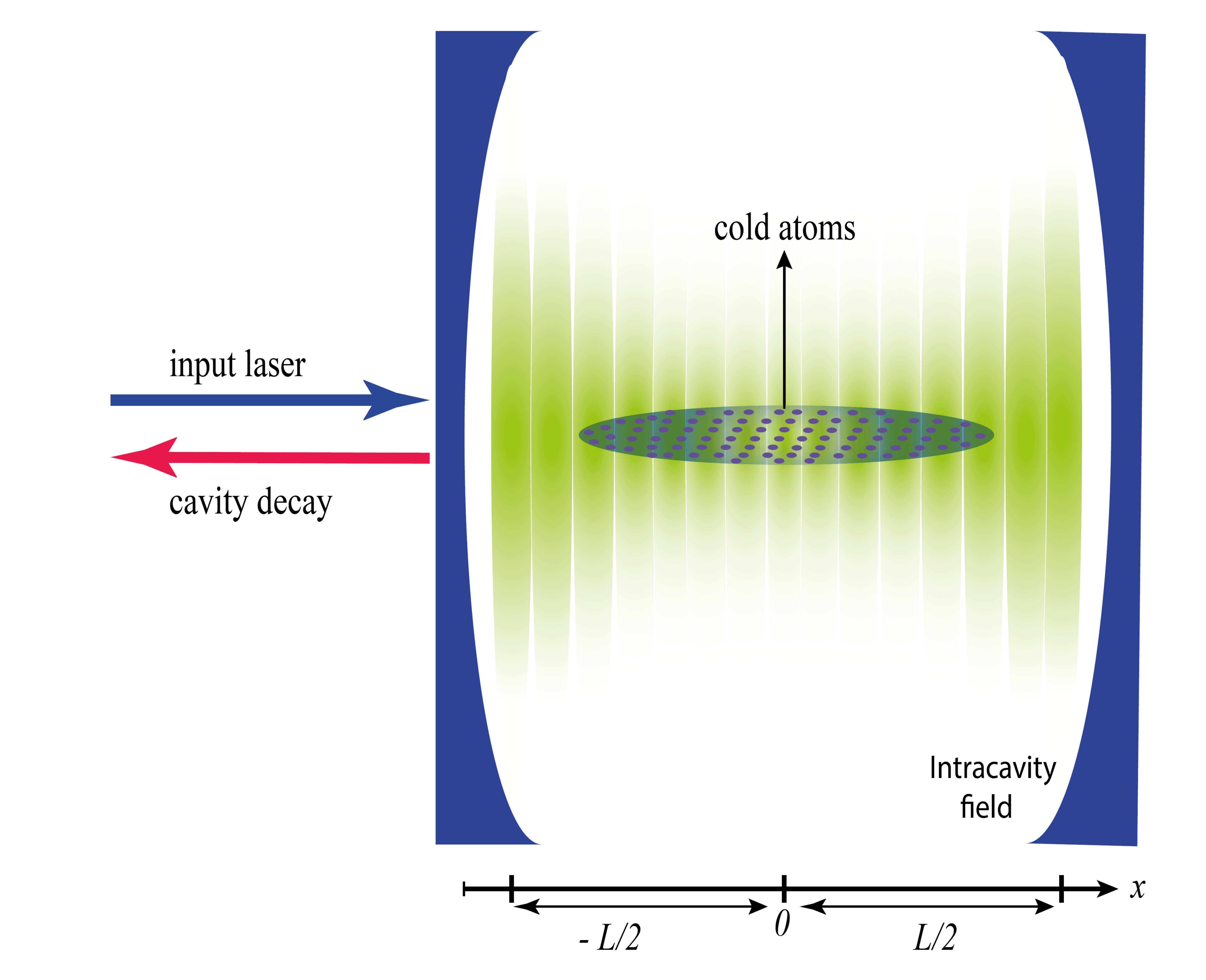}
\caption{
(Color online) N two-level atoms trapped in an optical cavity interacting dispersively with a single cavity mode. The cavity mode is driven by a laser at rate $\eta$ and the decay rate is $\kappa$.}
\label{fig:fig1}
\end{figure}

We are going to study a gas of $ N$ ultracold bosonic two-level atoms with mass $M$ and transition frequency $\omega_{a}$ inside the optical lattice of a single-mode, high-finesse Fabry-Perot cavity with length $L$. The cavity is driven at rate $\eta$ through one of its mirrors by a laser with frequency $\omega_{p}$, and wavenumber $K=2\pi/\lambda=\omega_{p}/c$. We assume the BEC is confined in a cylindrically symmetric trap with a transverse trapping frequency $\omega_{\mathrm{\perp}}$ and negligible longitudinal confinement along the $x$ direction (Fig.\ref{fig:fig1}). In this way we can describe the dynamics within an effective one-dimensional model by quantizing the atomic motional degree of freedom along the $x$ axis only.

\subsection{The General Form of The Hamiltonian of a BEC Inside an Optical Lattice}

In the dispersive regime where the laser pump is far detuned from the atomic resonance ($\Delta_{a}=\omega_{p}-\omega_{a}$  exceeds the atomic linewidth $\gamma$ by orders of magnitude), the excited electronic state of the atoms can be adiabatically eliminated and spontaneous emission can be neglected \cite{Masch Ritch 2004}. In the frame rotating at the pump frequency, the many-body Hamiltonian reads
\begin{equation}\label{H1}
H=-\hbar\Delta_{c} a^{\dagger} a + i\hbar\eta  (a^{\dagger}-a)+\int_{-L/2}^{L/2} dx \Psi^{\dagger}(x) H_{0} \Psi(x)+H_{aa},
\end{equation}
where $a$ is the annihilation operator for a cavity photon and $\Delta_{c}=\omega_{p}-\omega_{c}$ is the cavity-pump detuning. $H_{0}$ is the single-particle Hamiltonian of an atom inside the optical lattice of the cavity and $H_{aa}$ is the atom-atom interaction that are respectively given by
\begin{subequations}\label{hamil1}
\begin{eqnarray}
H_{0}&=&p^2/2M+\hbar U_{0} \cos^2(Kx) a^{\dagger} a,\\
H_{aa}&=&\frac{1}{2} U_{s} \int_{-L/2}^{L/2} dx \Psi^{\dagger}(x)\Psi^{\dagger}(x)\Psi(x)\Psi(x).
\end{eqnarray}
\end{subequations}
Here $U_{0}=g_{0}^{2}/\Delta_{a}$ is the optical lattice barrier height per photon and represents the atomic backaction on the field, $g_{0}$ is the vacuum Rabi frequency, $U_{s}=\frac{4\pi\hbar^{2} a_{s}}{M}$ and $a_{s}$ is the two-body s-wave scattering length \cite{Dom Ritch 2003,Masch Ritch 2005}. The second term in the Hamiltonian of Eq.(\ref{hamil1}a) is a periodic potential of period $\lambda/2$. 

It is well-known that the eigenfunctions of a particle inside a periodic potential are the Bloch functions $\psi_{\nu,q}(x)$ with eigenvalues $\epsilon_{\nu,q}$ where $\nu$ is the Bloch band index and $q\in[-2\pi/\lambda,2\pi/\lambda]$ is the quasimomentum of the particle \cite{Kittle}. If there are $l$ periods inside the cavity, then $L=l\lambda/2$. Using the Born-Von Karman periodic boundary condition $\psi_{\nu.q}(x+L)=\psi_{\nu,q}(x)$ and based on the Bloch theorem,
\begin{equation}\label{bloch1}
\psi_{\nu,q}(x+l\lambda/2)=e^{iq l\lambda/2}\psi_{\nu,q}(x),
\end{equation}
it is deduced that $q l\lambda/2=2m\pi$, where $(m\in \mathbb{Z})$. In this way the quasimomentum $q_{m}$ in the first Brillouin zone can be written in terms of optical wave number, i.e., $q_{m}=2mK/l$ where $-l/2\leqslant m \leqslant l/2$. Now using the Bloch theorem, the eigenfunctions $\psi_{\nu,q_{m}}(x)$ can be written as
\begin{equation}\label{bloch2}
\psi_{\nu,q}(x)=\frac{1}{\sqrt{l}} e^{iq_{m}x} u_{\nu}(x).
\end{equation}
Here $u_{\nu}(x)$ is a periodic function with the optical lattice period $(\lambda/2)$ and reads
\begin{equation}\label{u}
u_{\nu}(x)=\frac{1}{\sqrt{\lambda/2}} \sum_{n} c_{\nu,n} e^{i2nKx}.
\end{equation}
By substituting Eq.(\ref{u}) into Eq.(\ref{bloch2}) ,the eigenfunctions of the single-particle Hamiltonian $H_{0}$ is obtained as
\begin{equation}\label{blochf}
\psi_{\nu,q_{m}}(x)=\frac{1}{\sqrt{L}}\sum_{n} c_{\nu,n} e^{i(q_{m}+2nK)x}.
\end{equation}

The second quantized atomic wave field $\Psi(x)$ can be expanded as a Fourier series in terms of the Bloch functions, i.e.,
\begin{equation}\label{Psi}
\Psi(x)=\sum_{\nu,q_{m}}\tilde{b}_{\nu,q_{m}} \psi_{\nu,q_{m}}(x) .
\end{equation}
Here $\tilde{b}_{\nu,q_{m}}$ ($\tilde{b}_{\nu,q_{m}}^{\dagger}$) is the annihilation (creation) operator for the atomic field that annihilates (creates) a particle in a state determined with the Bloch band index $\nu$ and quasimomentum $q_{m}$. By using this expansion the Hamitlonian of the system is diagonalized and one can investigate the Bloch band structure of the condensate \cite{Bha2008}. However, the Fourier coefficients $ c_{\nu,n}$ can only be determined numerically and there is no closed form for the Bloch functions. Instead, by substituting Eq.(\ref{blochf}) into Eq.(\ref{Psi}) one can obtain a Fourier expansion of the atomic field in therms of plane waves
\begin{equation}\label{Psi2}
\Psi(x)=\frac{1}{\sqrt{L}}\sum_{n,m} b_{n,m} e^{i2K(n+m/l)x}.
\end{equation}
In the derivation of Eq.(\ref{Psi2}) we have done a Bogoliubov transformation
\begin{equation}\label{Bog}
b_{n,m}=\sum_{\nu}c_{\nu,n} \tilde{b}_{\nu,q_{m}},
\end{equation}
where $b_{n,m}$ ($b_{n,m}^{\dagger}$) is the atomic field operator that annihilates (creates) a particle in a state determined with the band index $n$ and quasimomentum $q_{m}$.

By substituting Eq.(\ref{Psi2}) into Eq.(\ref{H1}) the Hamiltonian of the systems can be written as
\begin{eqnarray}\label{H2}
H&=&-\hbar\Delta_{\mathrm{c}} a^{\dagger} a + i\hbar\eta  (a^{\dagger}-a)+\sum_{n,m}\varepsilon_{nm} b_{nm}^{\dagger} b_{nm}\nonumber\\
&&+ \frac{1}{4}\hbar U_{0} a^{\dagger} a\sum_{n,m} b_{nm}^{\dagger}(b_{n-1,m}+b_{n+1,m}+2b_{n,m})\nonumber\\
&&+H_{aa}.
\end{eqnarray}
In these sums $n\in \mathbb{Z}$, $-l/2\leqslant m \leqslant l/2$, and
\begin{equation}\label{epsilon}
\varepsilon_{n,m}=4E_{R}\Big(n+\frac{m}{l}\Big)^{2}
\end{equation}
are the energy eigenvalues in which $E_{R}=\hbar^{2}K^{2}/2M$  is the recoil energy. As seen, the field expansion of Eq.(\ref{Psi2}) can no longer diagonalize the Hamiltonian of the system. In fact that part of Hamiltonian which corresponds to the atom-photon interaction remains nondiagonal. Due to the interaction with the optical filed, the atoms are scattered to the nearest bands ($\Delta n=\pm1$) while their quasimomentum ($m$) remains unchanged.

\subsection{Optomechanical Model}
Since the optical potential is symmetric with respect to the origin, the Hamiltonian of the system has the parity symmetry. So if the sytem starts from a homogeneous BEC, only the cosine parts of the exponential functions in Eq.(\ref{Psi2}) are excited because of the parity conservation. In the limit of weak photon-atom coupling, when either the photon number inside the cavity or $U_{0}$ is small, the lowest bands $n=\pm1$ can be excited by fluctuations resulting from the atom-light interaction \cite{Nagy Ritsch 2009}. On the other hand, the s-wave scattering populates fluctuations with arbitrary $n$ and $m$ \cite{Szirmai 2010}. In a very simplified optomechanical model one can consider the scattering from $m=0$ to $m=\pm 1$ in the lowest band $n=0$. In this way the atomic field operator (Eq.(\ref{Psi2})) can be truncated as
\begin{equation}\label{opaf}
\Psi(x)=\frac{1}{\sqrt{L}}c_{00}+\sqrt{\frac{2}{L}} c_{10} \cos(2Kx)+\sqrt{\frac{2}{L}} c_{01} \cos(2Kx/l),
\end{equation}
where we have done the following Bogoliubov transformations
\begin{subequations}
\begin{eqnarray}
c_{nm}=\frac{1}{\sqrt{2}} (b_{nm}+b_{-n,-m}),\\
c_{n,-m}=\frac{1}{\sqrt{2}} (b_{n,-m}+b_{-n,m}).
\end{eqnarray}
\end{subequations}
In the case that the system does not have parity symmetry, for example when the BEC is inside a ring cavity, one should also consider sine modes with annihilation operators
\begin{equation}
s_{nm}=\frac{1}{\sqrt{2}} (b_{nm}-b_{-n,-m}),
\end{equation}
which in our model have been set aside \cite{Steinke Meustre 2011,Chen Meystre 2010}. By substituting the atomic field operator, Eq.(\ref{opaf}), into the Hamiltonian of Eq.(\ref{H1}), one can arrive at the following Hamiltonian
\begin{eqnarray}\label{OpA}
H&=&-\hbar \Delta_{c} a^{\dagger}a+i\hbar\eta (a-a^{\dagger})+4E_{R}(c_{10}^{\dagger}c_{10}+\frac{1}{l^2}c_{01}^{\dagger}c_{01})\nonumber\\
&&+H_{ac}+H_{aa},
\end{eqnarray}
where
\begin{subequations}\label{AC AA}
\begin{eqnarray}\label{abc}
H_{ac}&=&\frac{1}{2}\hbar U_{0}a^{\dagger}a(c_{00}^{\dagger}c_{00}+c_{10}^{\dagger}c_{10}+c_{01}^{\dagger}c_{01}\nonumber\\
&&+\frac{1}{\sqrt{2}}c_{00}^{\dagger}c_{10}+\frac{1}{\sqrt{2}}c_{00}c_{10}^{\dagger}),\\
H_{aa}&=&\hbar\frac{\omega_{sw}}{4N}(c^{\dagger 2}_{00}(c_{10}^{2}+c_{01}^{2})+c_{00}^{2}(c^{\dagger 2}_{10}+c_{01}^{\dagger2})\nonumber\\
&&+4c^{\dagger}_{00}c_{00}c^{\dagger}_{10}c_{10}\nonumber+4c^{\dagger}_{00}c_{00}c^{\dagger}_{01}c_{01}\nonumber\\
&&+c^{\dagger2}_{10}c_{01}^{2}+c^{2}_{10}c^{\dagger2}_{01}+4c^{\dagger}_{10}c_{10}c^{\dagger}_{01}c_{01}\nonumber\\
&&+\frac{3}{2}c^{\dagger2}_{01}c^{2}_{01}+\frac{3}{2}c^{\dagger 2}_{10}c_{10}^{2}+c^{\dagger 2}_{00}c_{00}^{2}).
\end{eqnarray}
\end{subequations}

The first two terms in Eq.(\ref{OpA}) denote the cavity and the pump Hamiltonians. The third term is the energy of the side-modes $c_{10}$ and $c_{01}$ which is just the third term in Eq.(\ref{H2}) with energy eigenvalues $\varepsilon_{nm}$ given by Eq.(\ref{epsilon}). The Hamiltonian $H_{ac}$ denotes the atom-photon interaction and $H_{aa}$ is the atom-atom interaction. Furthermore, $\omega_{sw}=8\pi\hbar a_{s}N/MLw^2$  is the s-wave scattering frequency and $w$ is the waist of the optical potential.
 
This Hamiltonian can be further simplified since for weak optical fields and large $N$ the depletion of the initial condensate remains weak. So we can use the Bogoliubov approximation and treat the zero-momentum mode classically, i.e., $c_{00}\rightarrow \sqrt{N}$ \cite{Kanamoto 2010, Goldbaum Meystre arXiv}. In this way Eqs.(\ref{OpA}) reduce to the following forms
\begin{subequations}\label{HB}
\begin{eqnarray}\label{subHB}
H&=&-\hbar\tilde{\Delta}_{c} a^{\dagger}a+i\hbar\eta (a-a^{\dagger})+\hbar(4\omega_{R}+\omega_{sw})c_{10}^{\dagger}c_{10}\nonumber\\
&&+\hbar(\frac{4}{l^2}\omega_{R}+\omega_{sw})c_{01}^{\dagger}c_{01}+H_{ac}+H_{aa},\\
H_{ac}&=&\frac{\sqrt{2N}}{4}\hbar U_{0} a^{\dagger}a (c_{10}+c^{\dagger}_{10})\nonumber\\
&&+\frac{1}{2}\hbar U_{0} a^{\dagger}a(c^{\dagger}_{10}c_{10}+c^{\dagger}_{01}c_{01}),\\
H_{aa}&=&\frac{1}{4}\hbar\omega_{sw}(c^{2}_{10}+c^{\dagger2}_{10}+c^{2}_{01}+c^{\dagger2}_{01}).
\end{eqnarray}
\end{subequations}
Here $\tilde{\Delta}_{c}=\Delta_{c}-NU_{0}/2$ is the effective Stark-shifted detuning. As seen from the Hamiltonian of Eq.(\ref{HB}b) there are two kinds of optomechanical coupling. The first term of Eq.(\ref{HB}b) is the linear radiation pressure which couples the side mode $c_{10}$ to the optical field with the optomechanical coupling constant $\frac{\sqrt{2N}}{4}\hbar U_{0}$ while the second term is the nonlinear optomechanical coupling of the two side modes $c_{10}$ and $c_{01}$ with the optical field. In the atom-atom interaction Hamiltonian $H_{aa}$ we have disregarded all terms proportional to $\omega_{sw}/N$.
The influence of s-wave scattering has partly appeared as a shift in the side modes energies (the third and fourth tems in Eq.(\ref{subHB})). Since $\hbar\omega_{sw}=2U_{s}n_{A}$, where $n_{A}=N/Lw^{2}$ is the density of atoms, it is apparent that the energy shits obtained here is the same as that of Bogoliubov theory \cite{Pethick}. On the other hand  the role of $H_{aa}$ in Eq.(\ref{HB}) for the Bogoliubov side modes is very similar to that of an optical parametric amplifier (OPA) in an optomechanical system with the nonlinear gain parameter $\omega_{sw}$ \cite{Agarwal 2009}. In Sec. V we will show how this parameter can affect the NMS  behaviour of the coupled Bogoliubov mode and the cavity field.
\section{Dyanamics of The Optomechanical System}
The dynamics of the optomechanical system described by Eqs.(\ref{HB}) is fully characterized by the following set of nonlinear QLEs , written in the frame rotating at the input laser frequency,
\begin{subequations}\label{NQL}
\begin{eqnarray}\label{subNQL}
\dot{a}&=&(i\tilde{\Delta}_{c}-\kappa)a-\frac{iU_{0}}{2}a[\frac{\sqrt{2N}}{2}(c_{10}+c^{\dagger}_{10})+c^{\dagger}_{10}c_{10}\nonumber\\
&&+c^{\dagger}_{01}c_{01}]-\eta+\xi,\\
\dot{c}_{10}&=&-(i\omega_{10}+\gamma)c_{10}-i\frac{\sqrt{2N}}{4}U_{0}a^{\dagger}a-\frac{iU_{0}}{2}a^{\dagger}a c_{10}\nonumber\\
&&-\frac{i}{2}\omega_{sw}c^{\dagger}_{10}+f_{10},\\
\dot{c}_{01}&=&-(i\omega_{01}+\gamma)c_{01}-\frac{iU_{0}}{2}a^{\dagger}a c_{01}-\frac{i}{2}\omega_{sw}c^{\dagger}_{01}+f_{01},\nonumber\\
\end{eqnarray}
\end{subequations}
where $\omega_{10}=4\omega_{R}+\omega_{sw}$ and $\omega_{01}=\frac{4}{l^2}\omega_{R}+\omega_{sw}$. Here $\kappa$ and $\gamma$ characterize the dissipation of the cavity field and collective density excitations of the BEC, respectively. The cavity-field quantum vacuum fluctuation $\xi(t)$ satisfies the Markovian correlation functions, i.e., $\langle\xi(t)\xi^{\dagger}(t)\rangle=(n_{ph}+1)\delta(t-t^{\prime})$, $\langle\xi^{\dagger}(t^{\prime})\xi(t))\rangle=n_{ph}\delta(t-t^{\prime})$ with the average thermal photon number $n_{ph}$ which is nearly zero at optical frequencies \cite{Gardiner}. Besides, $f_{10}(t)$ and $f_{01}(t)$ are the thermal noise inputs for the two side modes of BEC which also satisfy the same Markovian correlation functions as those of the optical noise. The noise sources are assumed uncorrelated for the different modes of both the matter and light fields.
\subsection{Linearization}
Now we are going to study the weak excitations of the condensate from its ground state. Such excitations include small deviations of both the atomic wave function and the optical field from their respective stationary states. So we decompose each operator in Eqs.(\ref{NQL}) as the sum of its steady-state value and a small fluctuation. By substituting $a=\alpha+\delta a$, $c_{10}=\sqrt{N}\beta_{1}+\delta c_{10}$ and $c_{01}=\sqrt{N}\beta_{0}+\delta c_{01}$ into Eqs.(\ref{NQL}) one can obtain a set of nonlinear algebraic equations for the steady-state values,
\begin{subequations}\label{ss}
\begin{eqnarray}
\alpha&=&\frac{\eta}{\sqrt{\Delta_{d}^2+\kappa^2}},\\
\beta_{1}&=&\frac{\sqrt{2}}{4}\frac{U_{0}\alpha^2}{\sqrt{\Omega^{(+)2}_{10}+\gamma^2}},\\
 \beta_{0}&=&0,
 \end{eqnarray}
\end{subequations}
where we have assumed $\alpha, \beta_{0}$ and $\beta_{1}$ are real numbers \cite{Konya Domokos 2011} and $\Delta_{d}=\tilde{\Delta}_{c}-\frac{1}{2}NU_{0}\beta_{1}(\beta_{1}+\sqrt{2})$ is the effective detuning,  $\Omega^{(\pm)}_{10}=\tilde{\omega}_{10}\pm\frac{1}{2}\omega_{sw}$ and $\tilde{\omega}_{10}=\omega_{10}+\frac{1}{2}U_{0}\alpha^{2}$. Eq.(\ref{ss}c) shows that the mean value of the mode $c_{01}$ is zero. It is the consequence of the fact that the ground state of the translationally invariant system is also invariant under discrete translation \cite{Szirmai 2010}. On the other hand, the linearized QLEs for the fluctuating operators take the followinf forms
\begin{subequations}\label{flc}
\begin{eqnarray}
\delta\dot{a}&=&(i\Delta_{d}-\kappa)\delta a-\frac{i}{2}G(\delta c^{\dagger}_{10}+\delta c_{10})+\xi,\\
\delta\dot{c}_{10}&=&-(i\tilde{\omega}_{10}+\gamma)\delta c_{10}-\frac{i}{2}G (\delta a+\delta a^{\dagger})\nonumber\\
&&-\frac{i}{2}\omega_{sw}\delta c^{\dagger}_{10}+f_{10},\\
\delta\dot{c}_{01}&=&-(i\tilde{\omega}_{01}+\gamma)\delta c_{01}-\frac{i}{2}\omega_{sw}\delta c^{\dagger}_{01}+f_{01},
\end{eqnarray}
\end{subequations}
where
\begin{equation}\label{G}
G=U_{0}\sqrt{N}\alpha(\beta_{1}+\frac{\sqrt{2}}{2})
\end{equation}
is the effective atom-photon coupling parameter. By defining the quadrature fluctuations
\begin{subequations}\label{flc}
\begin{eqnarray}
\delta X_{a}=\frac{1}{\sqrt{2}}(\delta a+\delta a^{\dagger}),  \delta P_{a}=\frac{1}{\sqrt{2}i}(\delta a-\delta a^{\dagger}),\\
\delta X_{j}=\frac{1}{\sqrt{2}}(\delta c_{j}+\delta c^{\dagger}_{j}),  \delta P_{j}=\frac{1}{\sqrt{2}i}(\delta c_{j}+\delta c^{\dagger}_{j}),
\end{eqnarray}
\end{subequations}
with $j=(10, 01)$, the QLEs can be written in the compact matrix form
\begin{equation}\label{nA}
\dot{u}(t)=A u(t)+n(t),
\end{equation}
where $u=[\delta X_{a},\delta P_{a},\delta X_{10},\delta P_{10},\delta X_{01},\delta P_{01}]^{T}$ is the vector of continuous variable fluctuation operators and
\begin{equation}
n(t)=[\xi_{x}(t),\xi_{p}(t),f_{x10}(t),f_{p10}(t),f_{x01}(t),f_{p01}(t)]^{T}.
\end{equation}
is the corresponding vector of noises. The $6\times6$ matrix $A$ is the drift matrix given by
\begin{equation}
A=\left(\begin{array}{cccccc}
-\kappa & -\Delta_{d} & 0 & 0 & 0 & 0 \\
   \Delta_{d} & -\kappa &-G &0 &0 & 0 \\
    0 & 0 & -\gamma & \Omega^{(-)}_{10} & 0& 0 \\
    -G & 0 & -\Omega^{(+)}_{10} & -\gamma & 0&0\\
  0& 0& 0 &0 &-\gamma & \Omega^{(-)}_{01} \\
   0 & 0 & 0 & 0 & -\Omega^{(+)}_{01} &-\gamma
  \end{array}\right).
\label{A}
\end{equation}
As is seen from the drift matrix, the side mode $(\delta X_{01},\delta P_{01})$ has been decoupled from the optical mode. It is due to the fact that in the QLEs, Eqs.(\ref{NQL}c), there is no standard radiation pressure coupling between the side mode $c_{01}$ and the optical field. Instead, it is coupled to the radiation field via the nonlinear term, $-\frac{iU_{0}}{2}a^{\dagger}a c_{01}$ which is as small as $1/\sqrt{N}$ of the radiation pressure term, and is deleted during the linearization process (because the mean field value of this mode, $\beta_{0}$, is zero). In this way the side mode $c_{01}$ acts as a medium level in the dynamics.
\subsection{Stationary Quantum Fuctuations}

In order to study the stationary properties of the system it is enough to focus our attention on the subspace spanned by the optical mode and the side mode $c_{10}$. It means that we can consider only the upper block of the drift matrix of Eq.(\ref{A}). The system is stable only if the real part of all the eigenvalues of the matrix $A$ are negative. These stability conditions can be obtained by using the Routh-Hurwitz criterion \cite{RH}. Due to the linearized dynamics of the fluctuations and since all noises are Gaussian the steady state is a zero-mean Gaussian state which is fully characterized by the $4\times4$ stationary correlation matrix (CM) $V$, with components $V_{ij}=\langle u_i(\infty)u_j(\infty)+u_j(\infty)u_i(\infty)\rangle/2 $. When the system is stable such a CM is given by\cite{Genes2008}
\begin{equation}\label{Vij}
V_{ij}=\sum_{k,l}\int_{0}^{\infty}ds\int_{0}^{\infty}ds'M_{ik}(s)M_{jl}(s')D_{kl}(s-s'),
\end{equation}
\begin{figure}[ht]
\centering
\includegraphics[width=3in]{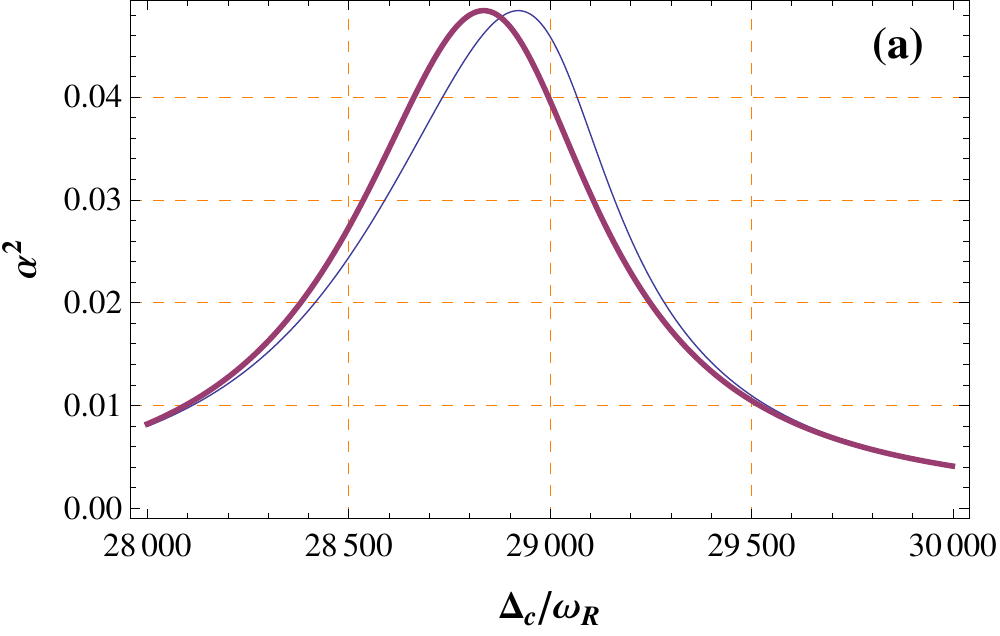}
\includegraphics[width=3in]{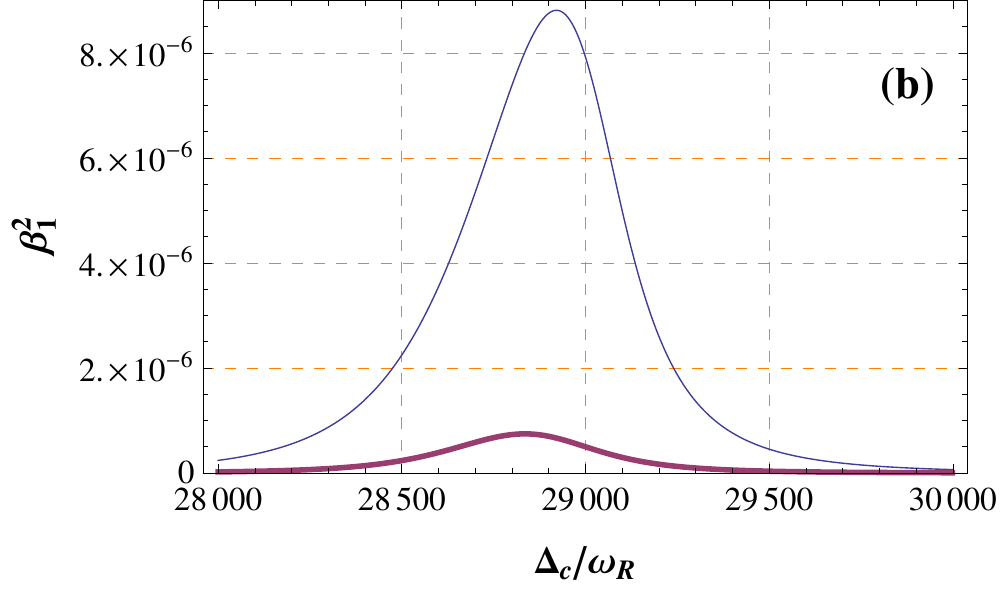}
\caption{
(Color online) (a) The mean cavity photon number and (b) the mean value fraction of atoms in the Bogoliubov side mode $c_{10}$ versus the normalized cavity detuning $\Delta_{c}/\omega_{R}$ for two values of $\omega_{sw}=\omega_{R}$ (thin line) and $\omega_{sw}=10 \omega_{R}$ (thick line). The parameters are $N = 6\times10^4$, $U_{0} = 0.96\omega_{R}$, $\kappa = 363.9   \omega_{R}$, $\gamma=0.001 \kappa$, $\eta=80.06\omega_{R}$ and $T=10^{-7} K$.}
\label{fig:fig2}
\end{figure}
\begin{figure}[ht]
\centering
\includegraphics[width=3in]{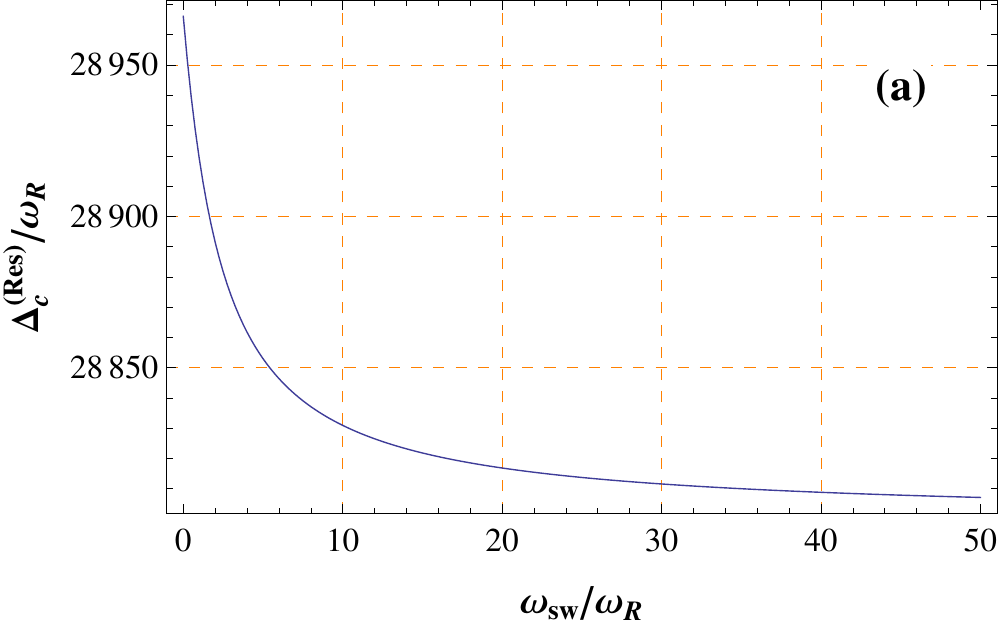}
\includegraphics[width=3in]{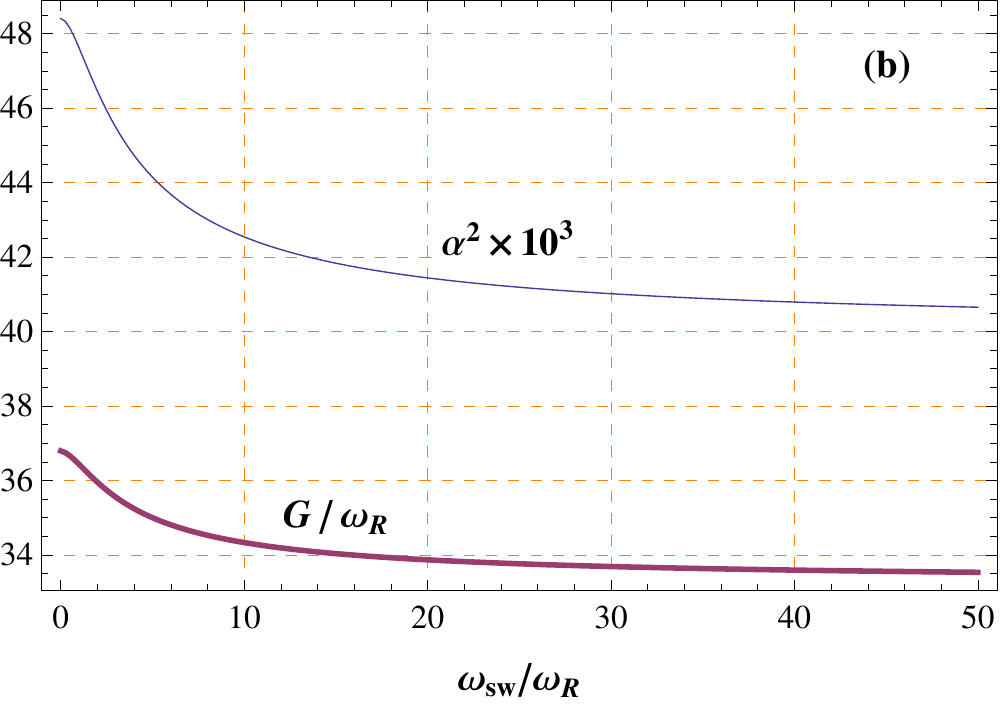}
\caption{
(Color online) (a) The normalized resonance frequency of the cavity $\Delta_{c}^{(Res)}/\omega_{R}$ and (b) the normalized effective atom-photon coupling $G/\omega_{R}$ (thick line) and the mean photon number magnified by $10^3$ (thin line) versus the normalized s-wave scattering frequency $\omega_{sw}/\omega_{R}$. The parameters are the same as those of Fig.\ref{fig:fig2}. The cavity detuning has been set at $\Delta_{c}=28966 \omega_{R}$.}
\label{fig:fig3}
\end{figure}
where $M(s)=\exp(As)$ and $D_{kl}(s-s')=\langle n_k(s)n_l(s')+n_l(s')n_k(s)\rangle /2 $ is the matrix of stationary noise correlation functions. For the noise diffusion matrix we have $D_{kl}(s-s')=D_{kl}\delta(s-s')$, where
$D_{kl}=\mathrm{Diag}[\kappa,\kappa,\gamma(2n_B+1),\gamma(2n_B+1)]
$ is the diffusion matrix, with $n_B=[\mathrm{exp}(\hbar \omega_m/k_B T)-1]^{-1}$ as the mean number of thermal excitations of the Bogoliubov side mode $(\delta X_{10},\delta P_{10})$ whose frequency of oscillations is given by 
\begin{equation}\label{wm}
\omega_{m}=\sqrt{\Omega^{(+)}_{10}\Omega^{(-)}_{10}}.
\end{equation}
Therefore, Eq.(\ref{Vij}) becomes
\begin{equation}\label{V2}
V=\int_{0}^{\infty}dsM(s)D M^T(s).
\end{equation}

When the stability conditions are satisfied $M(\infty)=0$ and Eq.~(\eqn{V2}) will be equivalent to the following Lyapunov equation for the steady-state CM
\begin{equation}\label{lyap}
AV+VA^T=-D.
\end{equation}
Equation(\ref{lyap}) is linear in $V$ and can be straightforwardly solved. However, the explicit form of $ V $ is complicate and is not reported here.

\section{Numerical Solutions to The Mean-Fields and Fluctuations}
In this section we first discuss our results based on the numerical solutions of Eqs.(\ref{ss}) for the mean fields and then solve Eq.(\ref{lyap}) to obtain fluctuations in the number of atoms and photons and their entanglement. We will show how the atom-atom interaction affects the mean fields, cavity resonance and atom-photon entanglement. We analyse our results based on the experimentally feasible parameters given in Ref.\cite{Brenn Science,Ritter Appl. Phys. B}.
\subsection{The mean-field solution}
Here we are going to study the effect of s-wave scattering on the behaviour of the mean-field values of the optical field and the Bogoliubov side mode $c_{10}$, given by Eqs.(\ref{ss}), quantitatively. We consider $N=6\times10^4$ atoms distributed in the optical cavity of length $L$=178 $\mu$m with bare frequency $\omega_{c}$ corresponding to a wavelength of $\lambda$=780 nm. The optical mode is coherently driven at rate $\eta$=80.06$\omega_{R}$ with the recoil frequency for rubidium atoms $\omega_{R}=2\pi\times$ 3.57 kHz and the one-atom light shift $U_{0}$=0.96 $\omega_{R}$ \cite{Brenn Science,Szirmai 2010}. 

In Fig.\ref{fig:fig2} we have plotted the mean number of photons and the fraction of condensate atoms occupying the Bogoliubov mode $c_{10}$ versus the normalized atomic detuning $\Delta_{c}/\omega_{R}$ at a fixed temperature of $T=10^{-7}$ K, and for two values of s-wave scattering frequencies $\omega_{sw}$. As is seen the increasing of the s-wave scattering interaction shifts the resonance frequency of the cavity to the lower values. Besides, it causes the number of atoms in the Bogoliubov mode to decrease i.e., leads to the depletion of this mode. This is in good accordance with the results obtained from numerical solution of the GPE (Fig.1 of Ref.\cite{Zhang 2009}). 

From Eq.(\ref{ss} a) one can obtain the resonance condition as
\begin{equation}\label{Res}
\Delta_{c}^{(Res)}=\frac{NU_{0}}{2}(1+\sqrt{2}\beta_{1}+\beta_{1}^{2}).
\end{equation}
Such a resonance shift has been obtained in Ref.\cite{Ritter Appl. Phys. B} without considering atomic collisions. As was mentioned in that reference the BEC acts as a Kerr medium that shifts the empty-cavity resonance. However, due to the atomic collisions the Bogoliubov side mode $c_{10}$ is depleted to other modes i.e., the mean value $\beta_{1}$ decreases. So a BEC with atom-atom interaction has a different resonance in comparison to a non-interacting one. For clarity, we have shown the effect of atom-atom interaction on the resonance of the cavity in Fig.\ref{fig:fig3}(a). For a non interacting BEC, i.e., $\omega_{sw}=0$, the resonance occurs at $\Delta_{c}^{(Rs)}$=28966 $\omega_{R}$ while for very large values of $\omega_{sw}$ it goes to $NU_{0}/2=28800 \omega_{R}$. In Fig.\ref{fig:fig3}(b) the effective atom-photon coupling parameter $G$ (thick line) , given by Eq.(\ref{G}), and the mean photon number of the cavity (thin line) have been plotted versus the normalized s-wave scattering frequency. This figure shows that both the effective atom-photon coupling parameter and the mean number of photons decrease with $\omega_{sw}$. All of these reductions are the direct consequence of the resonance shift exerted by atomic collisions. In the next subsection it will be shown that increasing the strength of atomic interaction causes the number of fluctuating photons to decrease. So the decrease of the total cavity photon number is a sign of atomic collisions. These results coincide with those obtained by numerical solution of GPE \cite{Horak 2000}.

\subsection{Fluctuations and Entanglement}

\begin{figure}[ht]
\centering
\includegraphics[width=3in]{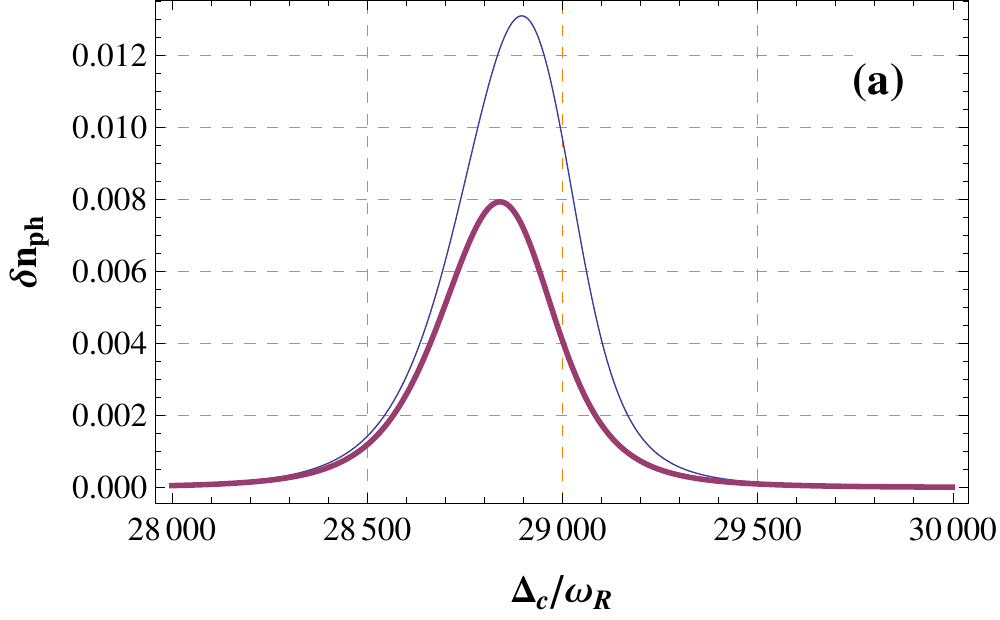}
\includegraphics[width=3in]{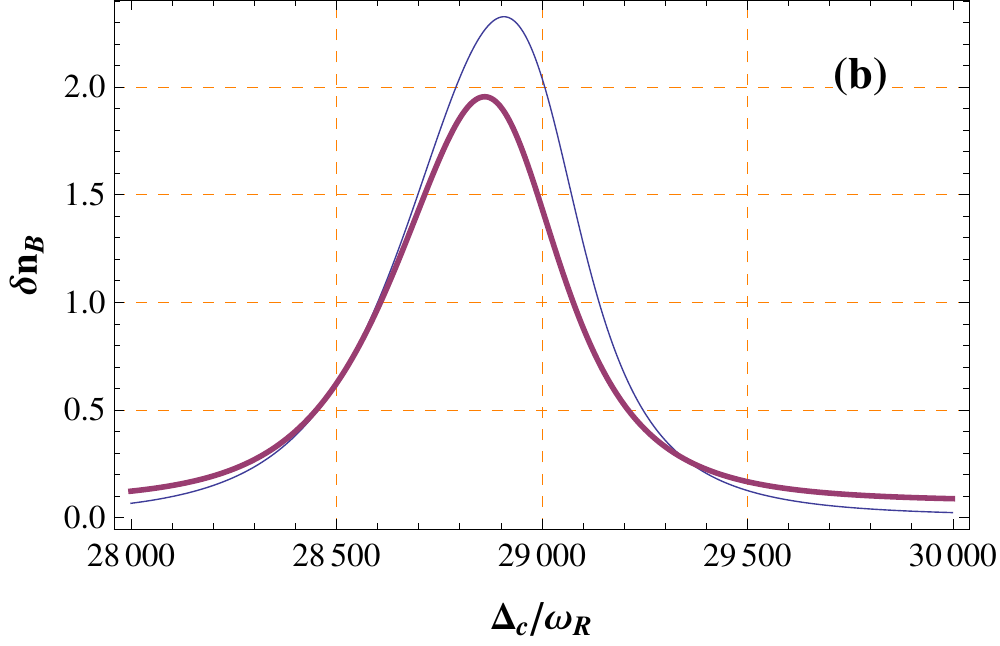}
\includegraphics[width=3in]{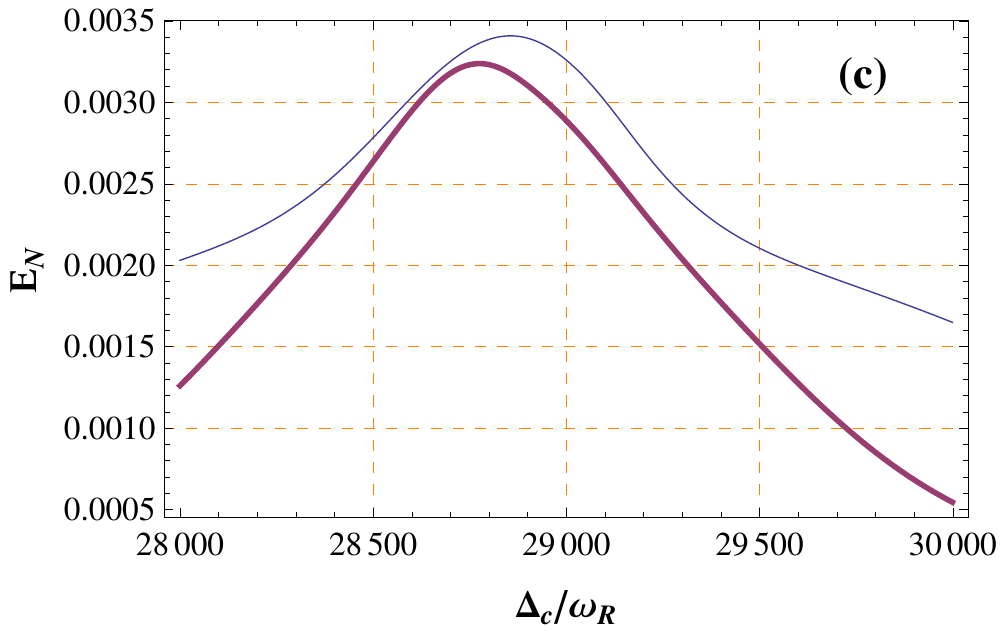}
\caption{
(Color online) The incoherent excitation numbers of the photons (a), the incoherent excitation numbers of atoms in the Bogoliubov side mode $c_{10}$ (b) and the the entanglement between the Bogoliubov mode $c_{10}$ and the optical field (c) versus normalized cavity detuning $\Delta_{c}/\omega_{R}$ for two different values of $\omega_{sw}=\omega_{R}$ (thin line) and $\omega_{sw}=10 \omega_{R}$ (thick line). All parameters are the same as those of Fig.\ref{fig:fig2}.}
\label{fig:fig4}
\end{figure}

After calculating the mean-field values we can obtain the elements of the drift matrix $A$ and solve for the steady solutions of Eq.(\ref{nA}). As explained before, by solving the Lyapunov equation (Eq.(\ref{lyap})) we can obtain the correlation matrix $V$ which gives us the second-order correlations of the fluctuations. The correlation matrix corresponding to the upper block of $A$ in Eq.(\ref{A}) can be written as
\begin{equation}\label{cV}
V=\left(
     \begin{array}{cc}
     \mathcal{A}&\mathcal{C} \\
      \mathcal{C}^{T} &\mathcal{B} \\
     \end{array}
   \right),
\end{equation}
where $\mathcal{A}$ and $\mathcal{B}$ represent the correlations of the photonic and atomic degrees of freedom respectively, and $\mathcal{C}$ describes the cross correlations. In this way we can calculate the incoherent excitation number of photons
\begin{equation}\label{deltanph}
\delta n_{ph}=\left\langle\delta a^{\dagger}\delta a\right\rangle=\frac{V_{11}+V_{22}-1}{2},
\end{equation}
and the incoherent excitation number of atoms in the Bogoliubov side mode
\begin{equation}\label{deltan10}
\delta n_{B}=\left\langle\delta c^{\dagger}_{10}\delta c_{10}\right\rangle=\frac{V_{33}+V_{44}-1}{2}.
\end{equation}
On the other hand the bipartite entanglement between the atomic and photonic degrees of freedom can also be calculated by using the logarithmic negativity \cite{eis}
\begin{equation}\label{en}
E_N=\mathrm{max}[0,-\mathrm{ln} 2 \eta^-],
\end{equation}
where  $\eta^{-}\equiv2^{-1/2}\left[\Sigma(V)-\sqrt{\Sigma(V)^2-4 \mathrm{det} V}\right]^{1/2}$   is the lowest symplectic eigenvalue of the partial transpose of the $4 \times 4$ CM, $V$,with $\Sigma(V)=\mathrm{det} \mathcal{A}+\mathrm{det} \mathcal{B}-2\mathrm{det} \mathcal{C}$.

\begin{figure}[ht]
\centering
\includegraphics[width=3in]{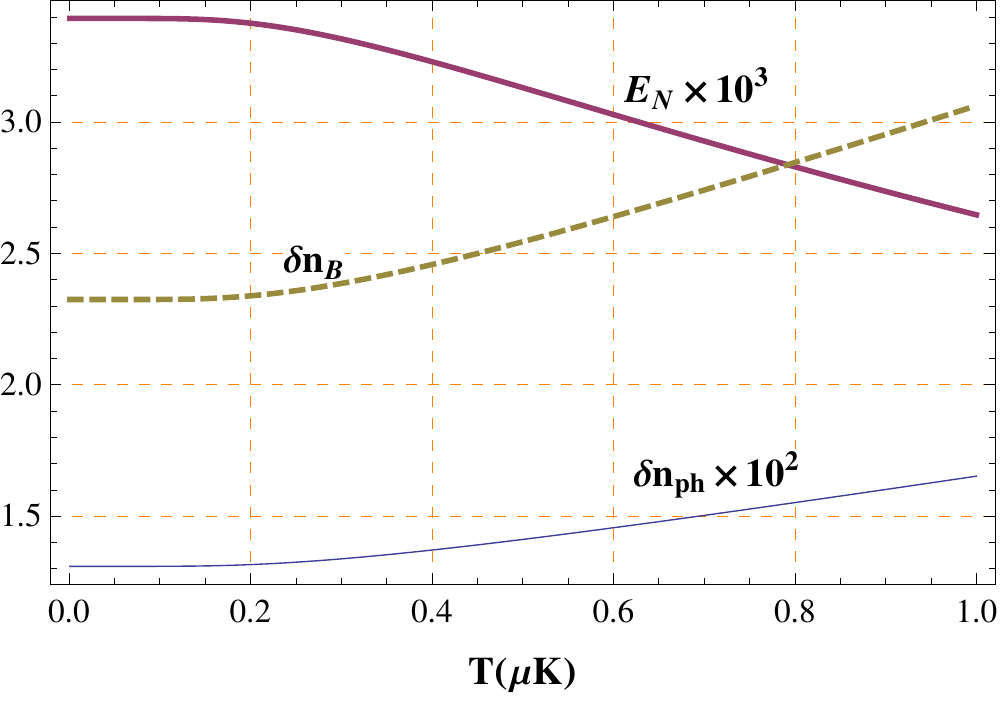}
\caption{(color online) The effects of temperature on the incoherent excitation numbers of photons (thin line) and atoms in the Bogoliubov side mode (dashed line) as well as the entanglement between photons and atoms (thick line). The cavity detuning has been set at $\Delta_{c}=28900 \omega_{R}$ and the s-wave scattering frequency has been considered to be $\omega_{sw}=\omega_{R}$. The other parameters are the same as those of Fig.\ref{fig:fig2}. For clarity, the values of $\delta n_{ph}$ and $E_{N}$ have been multiplied by 100 and 1000 respectively.}
\label{fig:fig5}
\end{figure}

In Figs.\ref{fig:fig4} (a) and (b) we have plotted, respectively, the incoherent excitation number of photons and atoms versus the normalized cavity detuning $\Delta_{c}/\omega_{R}$ for two values of $\omega_{sw}$. As is seen, fluctuations are small far from resonance but they increase rapidly near the resonance. On the other hand increasing atom-atom interaction decreases the fluctuations in the number of photons and atoms. Since the total number of photons is the mean value $\alpha^{2}$ that obtained in previous subsection plus the incoherent excitation number of photons, we can conclude that the stronger atom-atom interaction, the weaker optical field of the cavity. In fact the atom-atom interaction makes larger shifts in the cavity resonance frequency and consequently reduces the cavity field intensity. Hence the decrease of the cavity output provides a direct measure of the atom-atom interaction within the condensate.

in Fig.\ref{fig:fig4} (c) the entanglement between the Bogoliubov mode $c_{10}$ and the optical field has been plotted versus the normalized cavity detuning $\Delta_{c}/\omega_{R}$ for two values of $\omega_{sw}$. Again, the maximum of entanglement occurs at resonance. Besides, an increasing in the s-wave scattering frequency causes the entanglement of the atoms and photons to decrease. As it was mentioned above, an increase in atom-atom interaction shifts the resonance of the cavity which reduces the number of photons of the cavity and also causes the depletion of the Bogoliubov mode.This reduction in the mean values of photons and atoms leads to a reduction in the effective atom-photon coupling (Fig.\ref{fig:fig3}(b)) which causes the entanglement of atoms and photons to decrease.

The thermal effects on the incoherent excitation numbers of photons and atoms in the Bogoliubov side mode as well as the entanglement between photons and atoms have been illustrated in Fig.\ref{fig:fig5}. As is seen from the figure, increasing the temperature causes the fluctuations of the number of photons and atoms to increase and on the other hand reduces the atom-photon entanglement. Furthermore, the thermal effects for the range of temperatures below $0.1 \mu K$ are negligible. 

\section{Susceptibility and Power Spectrum of The Bogoliubov Mode}
Finally, we are going to obtain the power spectrum of the Bogoliubov side mode $c_{10}$ and also drive its effective frequency and damping rate. We will show that the coupling between the cavity field and the Bogoliubov mode which behaves like a mechanical mirror, leads to the splitting of the normal mode into two modes (NMS). A similar theoretical approach has been done in Ref.\cite{Bha 2010} for opto-mechanical Bose-Hubbard  Hamiltonian (OMBH) by expanding the atomic wave field in terms of the Wannier functions which is valid only for weak atom-field nonlinearity \cite{Larson 2008}. Instead, here we use the atomic field expansion, Eq.(\ref{opaf}), in the momentum space. For this purpose, we solve the linearized QLEs for the fluctuations in the displacement operator $\delta X_{10}$ as
\begin{equation}
\delta X_{10}(\omega)=\chi (\omega) F_{T}(\omega),
\end{equation}
where $F_{T}(\omega)$ is the Fourier transformation of the fluctuations in the total force acting on the Bogoliubov mode and $\chi(\omega)$ is its susceptibility which are given by, respectively,
\begin{subequations}
\begin{eqnarray}
F_{T}(\omega)&=&-\frac{G(\kappa-i\omega)}{\Delta_{d}^{2}+(\kappa-i\omega)^{2}}\xi_{x}(\omega)+\frac{G\Delta_{d}}{\Delta_{d}^{2}+(\kappa-i\omega)^{2}}\xi_{p}(\omega)\nonumber\\
&&+\frac{\gamma-i\omega}{\Omega_{10}^{(-)}}f_{x10}(\omega)+f_{p10}(\omega),\\
\chi(\omega)&=&\frac{\Omega_{10}^{(-)}}{\Omega_{eff}^{2}-\omega^{2}-i\omega\Gamma_{eff}},
\end{eqnarray}
\end{subequations}
where $\Omega_{eff}$ is the effective frequency of the Bogoliubov mode,
\begin{equation}\label{omegaeff}
\Omega_{eff}^{2}=\gamma^{2}+\omega_{m}^{2}+\frac{\Delta_{d}G^{2}\Omega^{(-)}_{10}(\Delta_{d}^{2}+\kappa^{2}-\omega^{2})}{(\Delta_{d}^{2}+\kappa^{2}-\omega^{2})^{2}+4\omega^{2}\kappa^{2}},
\end{equation}
and $\Gamma_{eff}$ is its effective damping rate,
\begin{equation}\label{gammaeff}
\Gamma_{eff}=2\gamma-\frac{2\kappa\Delta_{d}G^{2}\Omega^{(-)}_{10}}{(\Delta_{d}^{2}+\kappa^{2}-\omega^{2})^{2}+4\omega^{2}\kappa^{2}}.
\end{equation}
\begin{figure}[ht]
\centering
\includegraphics[width=3in]{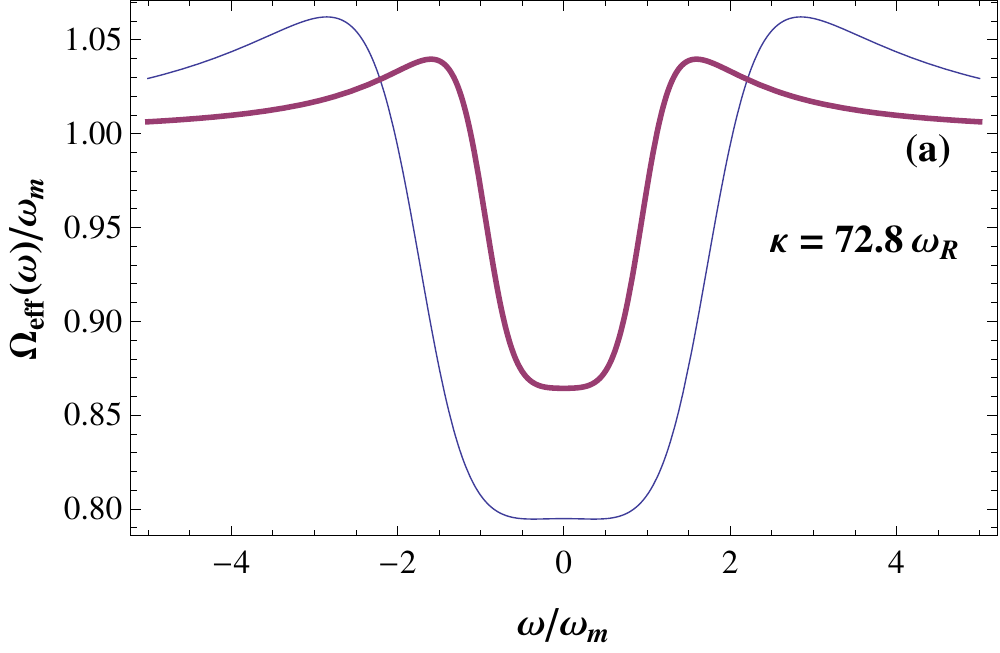}
\includegraphics[width=3in]{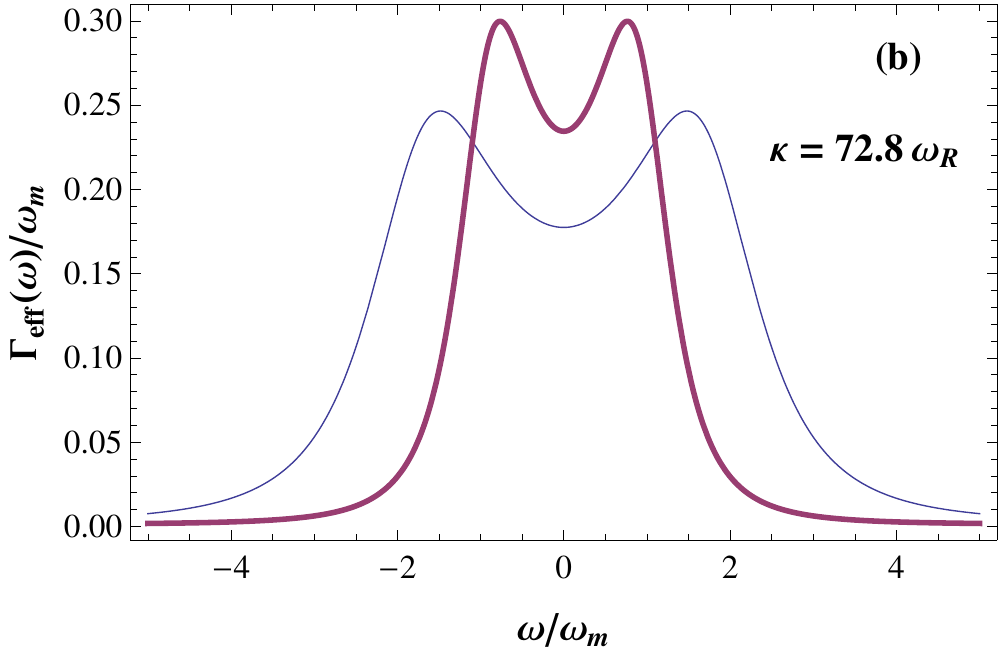}
\includegraphics[width=3in]{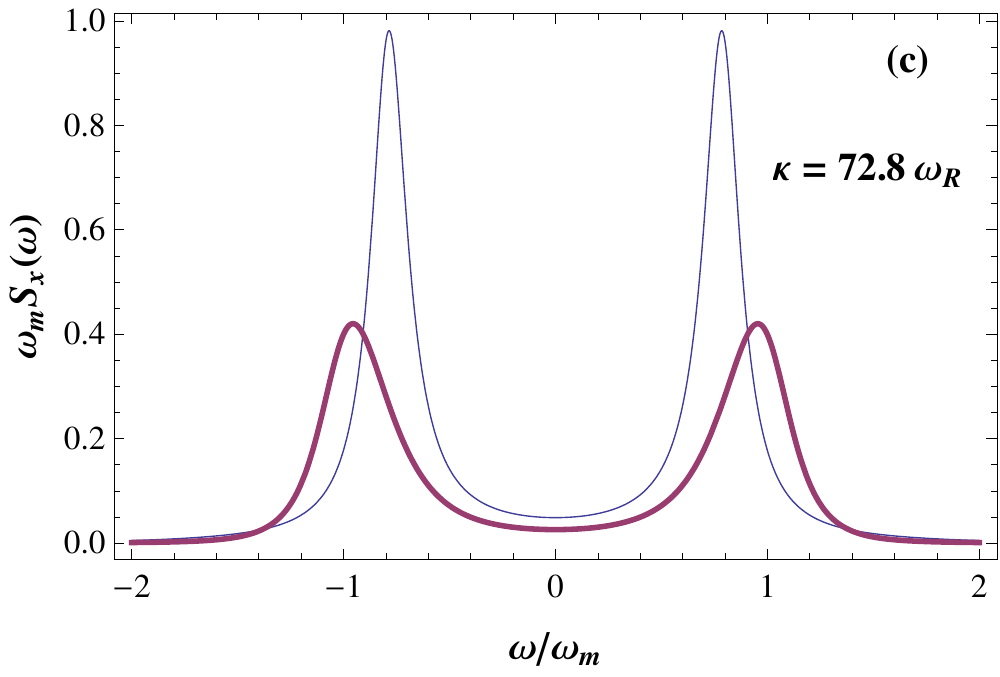}
\caption{ 
(Color online) (a) Normalized effective Bogoliubov mechanical frequency $\Omega_{eff}/\omega_{m}$, (b) normalized effective Bogoliubov mechanical damping $\Gamma_{eff}/\omega_{m}$ and (c) normalized power spectrum of the displacement operator of the Bogoliubov mode versus the normalized frequency $\omega/\omega_{m}$ for two values of $\omega_{sw}=80 \omega_{R}$ (thin line) and $\omega_{sw}=140 \omega_{R}$ (thick line) and for $\Delta_{c}=28700 \omega_{R}$ and $\kappa=72.8 \omega_{R}$. The other parameters are the same as those of Fig.\ref{fig:fig2}.}
\label{fig:fig6}
\end{figure}
\begin{figure}[ht]
\centering
\includegraphics[width=3in]{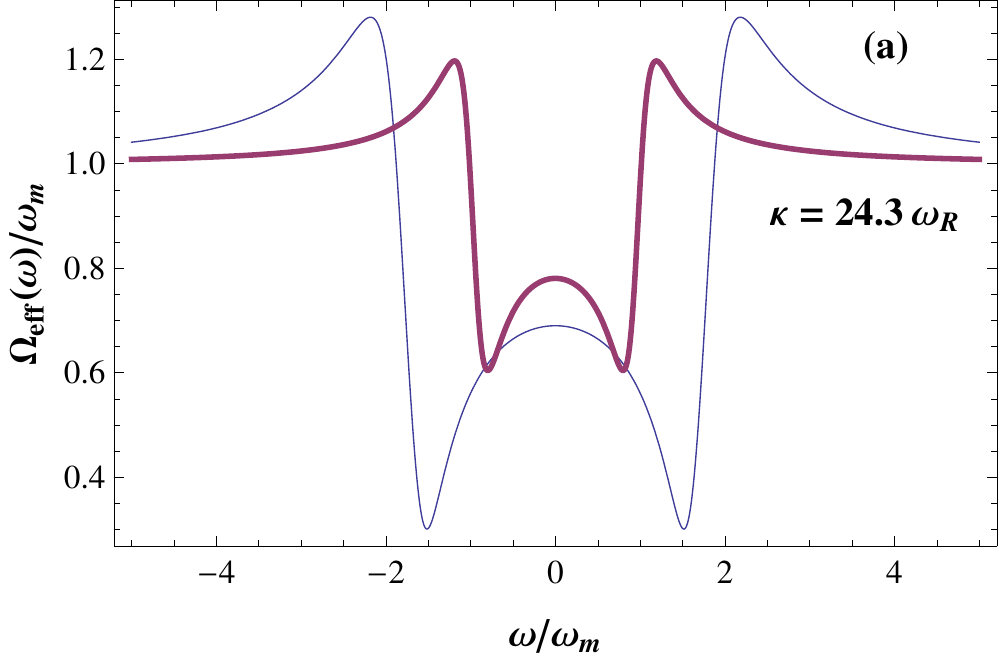}
\includegraphics[width=3in]{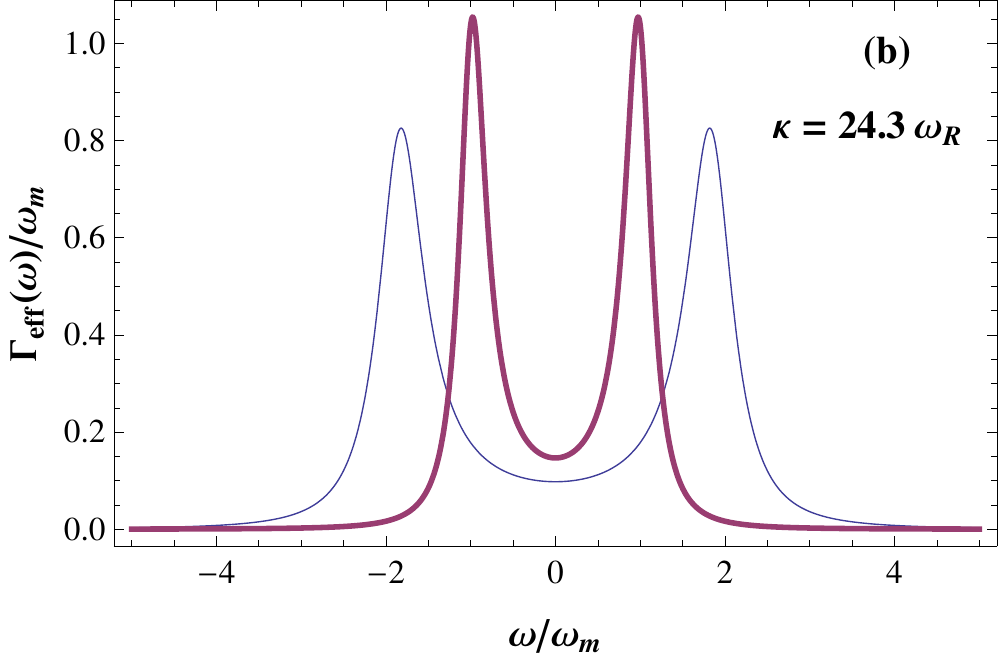}
\includegraphics[width=3in]{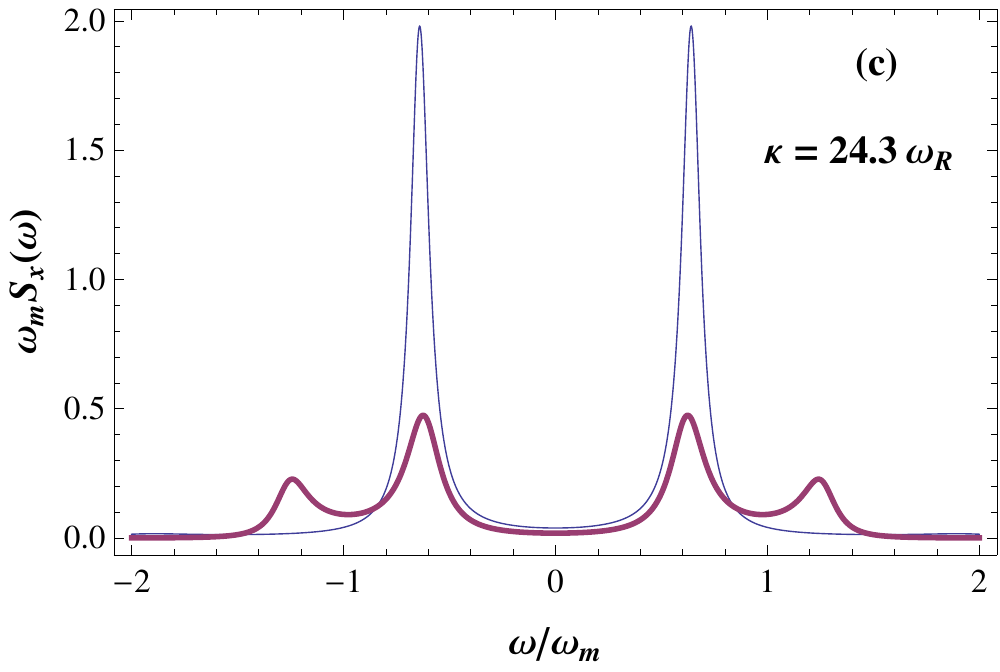}
\caption{ 
(Color online) (a) Normalized effective Bogoliubov mechanical frequency $\Omega_{eff}/\omega_{m}$, (b) normalized effective Bogoliubov mechanical damping $\Gamma_{eff}/\omega_{m}$ and (c) normalized power spectrum of the displacement operator of the Bogoliubov mode versus the normalized frequency $\omega/\omega_{m}$ for two values of $\omega_{sw}=80 \omega_{R}$ (thin line) and $\omega_{sw}=140 \omega_{R}$ (thick line) and for $\Delta_{c}=28700 \omega_{R}$ and $\kappa=24.3 \omega_{R}$. The other parameters are the same as those of Fig.\ref{fig:fig2}.}
\label{fig:fig7}
\end{figure}
To calculate the spectrum, we need the correlation functions of the optical noise sources in the frequency domain
\begin{subequations}
\begin{eqnarray}
\langle\xi_{x}(\omega)\xi_{x}(\omega^{\prime})\rangle=\langle\xi_{p}(\omega)\xi_{p}(\omega^{\prime})\rangle&=&\kappa \delta(\omega+\omega^{\prime}),\\
\langle\xi_{x}(\omega)\xi_{p}(\omega^{\prime})\rangle=\langle\xi_{p}(\omega)\xi_{x}(\omega^{\prime})\rangle^{\ast}&=&i\kappa \delta(\omega+\omega^{\prime}),
\end{eqnarray}
\end{subequations}
where we have assumed $n_{ph}=0$. For the atomic noise sources in the frequency domain we have
\begin{subequations}
\begin{eqnarray}
\langle f_{x}(\omega)f_{x}(\omega^{\prime})\rangle&=&2\gamma(n_{B}+\frac{1}{2}) \delta(\omega+\omega^{\prime}),\\
\langle f_{p}(\omega)f_{p}(\omega^{\prime})\rangle&=&2\gamma(n_{B}+\frac{1}{2}) \delta(\omega+\omega^{\prime}),\\
\langle f_{x}(\omega)f_{p}(\omega^{\prime})\rangle&=&i\gamma\delta(\omega+\omega^{\prime}),\\
\langle f_{p}(\omega)f_{x}(\omega^{\prime})\rangle&=&-i\gamma\delta(\omega+\omega^{\prime}),
\end{eqnarray}
\end{subequations}
where we have omitted the indices of $f_{x}$ and $f_{p}$ for simplicity. Here, $n_{B}=[\exp(\frac{-\hbar\omega_{m}}{k_{B}T})-1]^{-1}$ is the number of thermal excitations for the Bogoliubov mode and $\omega_{m}$ is the mechanical Bogoliubov frequency of oscillation given by Eq.(\ref{wm}). By using the relation $S_{x}(\omega)=\frac{1}{4\pi}\int d\omega^{\prime} e^{-i(\omega+\omega^{\prime})t}\langle\delta X_{10}(\omega)\delta X_{10}(\omega^{\prime})+\delta X_{10}(\omega^{\prime})\delta X_{10}(\omega)\rangle$ we can calculate the power spectrum of the displacement operator $\delta X_{10}(\omega)$ as follows
\begin{eqnarray}
S_{x}(\omega)&=&\frac{1}{4\pi}\vert\chi(\omega)\vert^{2} \Big[4\gamma (n_{B}+\frac{1}{2})\frac{\gamma^{2}+\omega^{2}+\Omega^{(-)2}_{10}}{\Omega^{(-)2}_{10}}\nonumber\\
&&+\frac{2\kappa G^{2}(\Delta_{d}^{2}+\omega^{2}+\kappa^{2})}{(\Delta_{d}^{2}+\kappa^{2}-\omega^{2})^{2}+4\omega^{2}\kappa^{2}}\Big].
\end{eqnarray}
The modification of the frequency of the Bogoliubov excitations of the condensate due to the radiation pressure given by Eq.(\ref{omegaeff}) is equivalent to the optical spring effect in cavity optomechanical systems with movable mirrors \cite{Genes 2008}.

In Figs. \ref{fig:fig6}(a) and \ref{fig:fig7}(a) we have plotted the normalized effective Bogoliubov mechanical frequency $\Omega_{eff}/\omega_{m}$ versus the normalized frequency $\omega/\omega_{m}$ for two values of cavity damping rates $\kappa=72.8 \omega_{R}$ (Fig.\ref{fig:fig6} (a)) and $\kappa=24.3 \omega_{R}$ (Fig.\ref{fig:fig7} (a)), and for two different values of $\omega_{sw}=80 \omega_{R}$ (thin line) and $\omega_{sw}=140 \omega_{R}$ (thick line). As is seen from both figures, a higher two-body interaction makes the condensate more robust and the Bogoliubov frequency of the condensate does not significantly deviate from $\omega_{m}$. Besides, a decrease in the damping rate of cavity causes the appearance of the peaks in the curves.

In Figs.\ref{fig:fig6}(b) and \ref{fig:fig7}(b) the normalized effective Bogoliubov mechanical damping $\Gamma_{eff}/\omega_{m}$ has been plotted versus the normalized frequency $\omega/\omega_{m}$ for two values of cavity damping rates $\kappa=72.8 \omega_{R}$ (Fig.\ref{fig:fig6} (b)) and $\kappa=24.3 \omega_{R}$ (Fig.\ref{fig:fig7} (b)). In the case of Fig.\ref{fig:fig6}(b) where the damping rate of the cavity is nearly equal to the effective atom-photon coupling constant $(\kappa\simeq G)$, the effective damping rate $\Gamma_{eff}$ is much lower compared to the case of Fig.\ref{fig:fig7}(b) where $G>\kappa$. It means that the stronger atom-photon coupling, the higher atom loss and hence the higher value of the effective damping. An experimental observation of this phenomenon of light induced back-action heating and consequent loss of atoms has been reported in Ref.\cite{Murch 2008}. It was also found that the atom loss rate is increased near the resonance. On the other hand, the effect of atomic collisions on the effective damping rate has been demonstrated in Figs.\ref{fig:fig6}(b) and \ref{fig:fig7}(b) which show that a stronger atom-atom interaction causes  the effective damping rate to be increased near the resonance points. Therefore increasing the rate of atomic collisions can help us in cooling of the Bogoliubov mode of the BEC by the radiation pressure.

Finally, we are going to investigate the influence of atomic collisions on the phenomenon of NMS in the power spectrum of Bogoliubov displacement operator. The NMS is associated with the mixing between the fluctuation of the cavity field around the steady state and the fluctuations of the condensate (Bogoliubov mode) around the mean field. The origin of the fluctuations of the cavity field is the beat of the pump photons with the photons scattered from the condensate atoms. The phenomenon of NMS is observable whenever the energy exchange between the two interacting modes takes place on a time scale faster than the decoherence of each mode.

In Figs.\ref{fig:fig6}(c) and \ref{fig:fig7}(c) we have shown the normalized power spectrum of the displacement operator of the Bogoliubov mode $S_{X}$ versus the normalized frequency $\omega/\omega_{m}$ for two values of cavity damping rate $\kappa=72.8 \omega_{R}$ (Fig.\ref{fig:fig6} (c)) and $\kappa=24.3 \omega_{R}$ (Fig.\ref{fig:fig7} (c)), and for two values of $\omega_{sw}=80 \omega_{R}$ (thin line) and $\omega_{sw}=140 \omega_{R}$ (thick line). In the case of Fig.\ref{fig:fig6}(c) where the damping rate of the cavity is nearly equal to the effective coupling constant of atoms and photons $(\kappa\simeq G)$, there is no splitting in the power spectrum (neither for $\omega_{sw}=80\omega_{R}$ nor for $\omega_{sw}=140\omega_{R}$) while in Fig.\ref{fig:fig7}(c) where $G>\kappa$ the NMS is appeared for $\omega_{sw}=140\omega_{R}.$ So in order to observe NMS firstly the system should be in the strong coupling regime where the atom-photon coupling is larger than the decay rates of photons an atoms. When this condition is fulfilled, the NMS can be observable with increasing atom-atom interaction. Normal mode splitting of a system of large number of atoms coupled to a ring cavity has been observed experimentally in the strong cooperative coupling regime \cite{Klinner}. The splitting of the normal mode has been observed by increasing the number of atoms which leads to the increase in the s-wave scattering frequency.

\section{Conclusions}
In conclusion, we have done a theoretical investigation on the optomechanical properties of a  one-dimensional Bose-Einstein condensate inside a driven optical cavity considering the effects of atomic collisions. Due to the dispersive atom-photon interaction the atoms develop a band structure in the optical lattice of the cavity. In the limit of weak photon-atom coupling the lowest bands $n=\pm1$ can be excited by fluctuations due to the atom-light interaction.  On the other hand, the atomic collisions scatter the atoms to the higher band and also populate states with non zero quasimomentum. It has been shown that there is a nonlinear optomechanical coupling between the nonzero quasimomentum states with the optical field proportional to $1/\sqrt{N}$ which can be disregarded in the Bogoliubov approximation where the number of atoms in the BEC mode is very large. Therefore, there is no radiation pressure coupling between the atomic modes with nonzero quasimomentom and the cavity field in the Bogoliubov approximation. 

In this way we have obtained a simplified optomechanical model where the Bogoliubov side mode $c_{10}$ is coupled to the optical field through the radiation pressure term and the atom-atom interaction Hamiltonian behaves like an OPA for the Bogoliubov mode with the non-linear gain parameter $\omega_{sw}$. It has been shown that in the strong coupling regime where the effective atom-photon coupling is greater than the cavity damping rate, an increase in the s-wave scattering frequency would lead to the NMS in the power spectrum of the Bogoliubov mode.

Furthermore, the atom-atom interaction causes the depletion of the Bogoliubov mode and also shifts the cavity resonance frequency which leads to a decrease in the mean number of cavity photons. Besides, it decreases fluctuations in the number of photons and atoms. Hence the decrease in the cavity output provides a direct measure of the atom-atom interaction within the condensate. On the other hand, a stronger atomic collision rate causes a decrease in the effective atom-photon coupling which leads to a decrease in the entanglement between the Bogoliubv mode and the optical field.

\section*{Acknowledgement}
The authors wish to thank The Office of Graduate
Studies of The University of Isfahan for their support.

\bibliographystyle{apsrev4-1}

\end{document}